\def\year{2019}\relax
\documentclass[letterpaper]{article} 
\usepackage{amsmath,amssymb,amsfonts}
\usepackage{algorithmic}
\usepackage{graphicx}
\usepackage{textcomp}
\usepackage{xcolor}
\def\BibTeX{{\rm B\kern-.05em{\sc i\kern-.025em b}\kern-.08em
    T\kern-.1667em\lower.7ex\hbox{E}\kern-.125emX}}
\usepackage{dirtytalk}
\usepackage{aaai19}  
\usepackage{times}  
\usepackage{helvet} 
\usepackage{courier}  
\usepackage[hyphens]{url}  
\usepackage{graphicx} 
\usepackage{graphicx}  
\nocopyright
\title{Trust and Cognitive Load During Human-Robot Interaction}
\author{Muneeb Imtiaz Ahmad\textsuperscript{\rm 1}, Jasmin Bernotat\textsuperscript{\rm 2}, Katrin Lohan\textsuperscript{\rm 3}, Friederike Eyssel\textsuperscript{\rm 4}\\ 
\textsuperscript{\rm 1,3}Edinburgh Center For Robotics, Heriot-Watt University, Edinburgh, United Kingdom\\
\textsuperscript{\rm 2,4}CITEC, Bielefeld University, Bielefeld, Germany\\
\textsuperscript{\rm 1}m.ahmad@hw.ac.uk,
\textsuperscript{\rm 2}jbernotat@techfak.uni-bielefeld.de
\textsuperscript{\rm 3}k.lohan@hw.ac.uk, 
\textsuperscript{\rm 4}friederike.eyssel@uni-bielefeld.de\\
}
 
\begin{document}

\maketitle

\begin{abstract}
This paper presents an exploratory study to understand the relationship between a humans' cognitive load, trust, and anthropomorphism during human-robot interaction. To understand the relationship, we created a \say{Matching the Pair} game that participants could play collaboratively with one of two robot types, Husky or Pepper. The goal was to understand if humans would trust the robot as a teammate while being in the game-playing situation that demanded a high level of cognitive load. Using a humanoid vs. a technical robot, we also investigated the impact of physical anthropomorphism and we furthermore tested the impact of robot error rate on subsequent judgments and behavior. Our results showed that there was an inversely proportional relationship between trust and cognitive load, suggesting that as the amount of cognitive load increased in the participants, their ratings of trust decreased. We also found a triple interaction impact between robot-type, error-rate and participant's ratings of trust. We found that participants perceived Pepper to be more trustworthy in comparison with the Husky robot after playing the game with both robots under high error-rate condition. On the contrary, Husky was perceived as more trustworthy than Pepper when it was depicted as featuring a low error-rate. Our results are interesting and call further investigation of the impact of physical anthropomorphism in combination with variable error-rates of the robot.
\end{abstract}

\section{Introduction}

The deployment of robotics systems can be witnessed in offshore environments \cite{hastie2019challenges}.  Under these settings, human operators are expected to interact with robots and also maintain a level of situation awareness allowing them to successfully supervise the autonomous operations \cite{lopes2019towards}. It is speculated that under emergencies in these environments, operators may experience a high amount of cognitive load and this may adversely impact their trust in the system. We, therefore, are particularly interested in the investigation of the relationship between cognitive load and trust and also other factors that could lead to maximizing users' trust perception during Human-Robot Interaction (HRI). We believe through understanding about these factors, we can create robotic systems that can minimize user's cognitive load and maximize user's trust.



Most recently, a meta-analysis on the factors impacting users' trust during HRI revealed that the robot's task performance and attributes such as anthropomorphism and proximity are among significant factors that result in enhanced user trust \cite{b3}. However, it should also be noted that this meta-analysis 
was based on a limited number of studies and also presented differing results on the factors impacting user trust. Most recently, Bernotat et al. (2018)  investigated German and Japanese judgments of an anthropomorphic vs. an industrial robot in smart homes. They found a marginally significant difference in Japanese vs. German participants' trust ratings, but they did not find an impact of robot type on trust. That is, participants indicated having the same level of trust in an anthropomorphic robot as they did in an industrial robot. Trust ratings were generally rather low in both cases \cite{bernotat2018can}. Anthropomorphism has also been highlighted as a critical factor in the literature on human-automation use \cite{b5}. It was observed in 
one of the studies that participants significantly trusted a machine more with the anthropomorphic condition in comparison to the mere \textit{Agentic} condition \cite{b4}. In summary, existing literature suggests that anthropomorphism impacts humans trust. In addition, we also find limited research on understanding the impact of anthropomorphism on trust in HRI and recognize the need for investigating this impact during future studies.



Users' cognitive load has also been identified as one of the significant factors impacting users' trust perception toward system during Human-Machine Interaction (HMI). It has been observed that both trust and cognitive load mediate user behavior during HMI \cite{b6}. This suggests that, if a user 
experiences a high amount of cognitive load, this may result in a degradation of their task performance and may also lower their trust perception. 
More importantly, it has also been demonstrated that participants' subjective rating of cognitive load decreases as the autonomy increases during teleoperated robot scenarios \cite{draper1996workload}. 
As one of 
our goals 
is to maximize operator's trust in the robotic system in the offshore environments, and it is expected that these environments may demand higher cognitive load, therefore, we are interested in understanding the relationship between trust and situations demanding high amount of cognitive load.
Despite the afore-mentioned significance of the relationship between cognitive load and trust perception during HMI, we, to the best of our knowledge, find very little research on investigating the relationship between user's trust and cognitive load in HRI and believe that the current research can crucially contribute to closing this research gap and also help us in achieving our end goals. 
We, therefore, keeping these aforementioned factors in mind, 
present a study to investigate the relationship between the concepts of Trust, Cognitive load and Anthropomorphism. In particular, our motivation is to understand how robots can gain trust from humans to ensure a smooth interaction and this may result in reducing cognitive load during HRI. 

To achieve our goals, we programmed a \say{Matching the Pairs} game and adapted it to the Husky\footnote{https://clearpathrobotics.com/husky-unmanned-ground-vehicle-robot/} and the Pepper\footnote{https://www.softbankrobotics.com/emea/en/pepper} robots. The rationale for choosing the Husky and the Pepper was to compare non-humanoid-like and humanoid-like robots. The goal is to enable both robots to play the game as teammates with human users. More specifically, we are interested in exploring the concept of ‘teammates’ in relation to the HRI during the \say{Matching the Pairs} game. We intend to determine whether humans trust the robot to undertake its assigned tasks, whether the robot can assist  humans in achieving their goal and whether the establishment of a Human-Robot team is possible. Additionally, we also intend to understand the relationship between cognitive load and trust during such robotic interactions. We achieve this through examining whether the participants trust the robot as a teammate in guiding them and facilitating in the selection process to achieve the overall goal of the game (finding the other matching pair). To induce cognitive load, subjects are expected to use their memory to recall positions of the matching pair, while at the same time communicating with the robot and striving for either a good score or just to complete the \say{Matching the Pair} game. This setup was considered to induce high cognitive load for the participants, thus, making it a good testing ground to investigate the relationship between the cognitive load and trust perception of the users. 

\section{Background}

\subsection{Trust and Factors impacting Trust in HRI}

In general, trust is considered to be a crucial aspect for a team functioning and collaborating successfully. Each member of the team must establish trust, apply it in their work, share it and maintain it. As humans are expected to collaborate with robots in various settings in the future, trust is also one of the critical measures to ensure successful HRI during various social settings. For instance, Hancock et al. (2011) identified trust as an important factor to consider, as the presence or absence of trust directly impacts the outcome of HRI \cite{b3}. 

As discussed above, trust in HRI is a complex concept dependent on a number of factors. These factors are divided into three different categories: robot-related, human-related, and environmental characteristics. Robot-related factors including attributes such as anthropomorphism, robot predictability, and robot personality, have a large part to play in the success of the HRI. Similarly, human-related characteristics play a major role in the capability of the human to trust the robot \cite{b2,schaefer2013perception}. For example, a human's understanding of the capabilities of the robot may have been shaped by the portrayals of robots in TV and film; this, in turn, would result in either an exaggerated or understated expectation of the capabilities of the robot. 
A recent study found that most participants indicated to know robots only from the media. It is thus likely that their perception of robots is influenced by the image media creates about robots \cite{bernotat2019fe}. 
Lastly, environmental factors impacting trust would be if a human is trained in close proximity to a robot, it encourages trust in the robot. More specifically, Lee \& See (2004)
investigated trust in interpersonal relationships and determined a trust importance scale based on the type of environment; trust is less important in a fixed, well-structured environment, for example, procedural based hierarchical organizations in which the order and stability decrease the uncertainty \cite{b7}. In contrast, trust is important within dynamic environments, for example, the interaction game for this paper \say{Matching the Pairs} where the game environment is dynamic, and the participant is uncertain of which marker to choose to make a match. In addition to the aforementioned factors, researchers have observed several other factors including impacts of culture, the type of task assigned, task complexity, the provision of training, and the amount of risk involved in the task and in the interaction between human and robot \cite{b3}. 

Of the three antecedents of trust proposed by \cite{b3}, robot characteristics were acknowledged as the most important factors to consider in trust development. Environmental characteristics provide a neutral outcome on trust, whilst there is some debate on the degree of importance of human characteristics on trust \cite{b3}. Consequently, we focus on the aspects of robot-related characteristics and their impact on trust in HRI.

\subsection{Trust and Cognitive Load}

Cognitive load refers to the amount of effort placed on the working memory during a task. According to Sweller \cite{b10}, there are three different types of cognitive load produced during problem-solving or learning: 1) intrinsic load, 2) the extraneous load and 3) the germane load. Intrinsic load depends on the complexity of the structure of the material and its association with the learner,
the extraneous load is caused by the way this material is presented to the learner, while the germane load is a result of the learner's ability to assimilate the material \cite{b10}. 

In the past, researchers have observed that both trust and cognitive load are related to each other. They have observed that as cognitive load increases, it will negatively impact human's trust perception \cite{b6}. As highlighted, we find limited research on the investigation of the relationship of trust and cognitive load in the HRI domain, therefore, we try to understand this relationship from the literature on human-automation/machine use. The common finding of the studies conducted in Human-Automation Interaction suggests that users prefer to work with the pre-existing trusted system instead of a new system under higher cognitive load \cite{b11}\cite{b12}. In particular, Biros et al. \cite{b12} conducted a study to investigate the variation of human trust in the automated system during a situation involving higher cognitive load. Their results showed that under high cognitive load situations, humans continue to rely on the existing system, although they have less trust in the system. In another study, it was also found that humans show an enhanced level of trust in the system that maintains its reliability and dependability \cite{b13}. Similarly, it has also been shown in one of the studies that humans prefer to use automation during increased workload situations \cite{b8}. 
On the contrary, one study 
also found that humans trust perception measured subjectively about automation decreases during higher workload situations as they prefer manual work over automation in such situations \cite{b9}. However, it has been argued that trust measured subjectively on the automation use is not a correct reflection of the relationship between trust and cognitive load \cite{b8}. Additionally, this may have resulted due to an automation bias of the participants. A recent review suggested that this variation may have happened due to cultural differences \cite{chien2018effect}. Conclusively, most of the past studies \cite{mcbride2011understanding,wang2011effects}
suggests that the decrease in the amount of humans' trust perception of the automated system may increase the level of humans' cognitive load. Also both trust and cognitive load are closely related and they impact human behavior. More specifically, if humans trust, they may stop weighing up pros and cons of working with robots, thus, freeing up cognitive load. 


To summarize, we believe that we find a number of studies exploring the relationship between trust and cognitive load in human-automation use. However, there is a limited exploration of this relationship in HRI. The findings from the Human-Automation Interaction studies may also not be applied to HRI. Hence, this calls for the deeper investigation of the impact of cognitive load and trust during HRI. 
Therefore, we investigated 
the relationship between cognitive load and trust during the Human-Robot game (\say{Matching the Pairs}) interaction.


\subsection{Measuring Cognitive Load and Trust}

Past research has identified a number of factors observable as a consequence of high cognitive load. These factors were commonly based on the human's linguistic behavior. These behaviors included: enhanced use of pausing, hesitations, and self-corrections \cite{b14,b15,b16,lopes2018symptoms}, enhanced use of negative emotions, decreased use of positive emotions and several other indicators \cite{b17,b18}. We, on the contrary present an approach called \say{Pupillometry} to measure cognitive load during HRI.

Pupillometry is used as a term in psychology and neurology to describe the act of measuring the diameter of a person's pupil 
It has been established that changes in one's eyes' pupil diameter is an indicator of cognitive activity \cite{hess1960pupil,hess1975tell}. For instance, German neurologist Bumke had already recognized at the beginning of the previous century that every intellectual or physical activity translates into pupil enlargement \cite{hess1975tell}. 
Hess and Polt (1960) achieved a major milestone for pupillometry by discovering that showing semi-nude photos of adults to subjects of the opposite sex would cause their pupils to dilate twenty percent on average \cite{hess1960pupil}. This study provided concrete proof that emotional stimulation causes enlargement of pupil diameter. This notion was later expanded to include other cognitive processes such as memory and problem-solving. Beatty and Kahneman (1966) showed that storing an increasing number of digits in one's memory would cause pupillary dilation \cite{beatty1966pupillary}, while it was also shown by Hess and Polt (1964) that pupil size corresponds with the difficulty of a cognitive task \cite{hess1964pupil}.

More recently Just and Carpenter (1993) showcased that pupil responses can be an indicator of the effort to comprehend and process information. They conducted an experiment where participants were given two sentences of different complexities to read while they would measure their pupil diameters. Pupillary dilation was larger while readers processed the sentence that was deemed to be more complicated and more subtle while they read the simpler one \cite{just1993intensity}. We believe that these findings make the connection between 
pupillometry and cognitive load clear, as they demonstrate that changes in the properties of an element to be processed (e.g. changing the complexity of a sentence in turn impacting the amount of intrinsic load it will impose on the reader) causes different pupillary responses. Therefore, we present pupillometry as a measure of cognitive load during HRI.

To measure pupillometry during HRI, we used Tobii Glasses Pro 2 eye tracking glasses\footnote{https://www.tobiipro.com/product-listing/tobii-pro-glasses-2/}, to estimate amounts of cognitive load
experienced by the user. 


We used a Godspeed questionnaire \cite{b19}, containing additional questions related to the experiment to measure trust. The additional questions were our primary measure of trust and the Godspeed was used to compare how participants felt about the robot before and after exposure. 
It is important to note that all Godspeed items were used but only items on trust were d to study the relationship between trust and cognitive load as the other items were beyond the scope of our analysis. 

\section{System Description}

Our system comprised of a "Matching the pair game" and we enabled both the Husky and the Pepper to play the game with the users through playing the role of a teammate. The concept is to explore the element of trust in the robot shown by the user during the game. We also measured and recorded the level of human cognitive loads. 
This was done to understand the relationship between users' cognitive load and their trust perception. 

\begin{figure}
    \centering
    \includegraphics[width=0.5\textwidth]{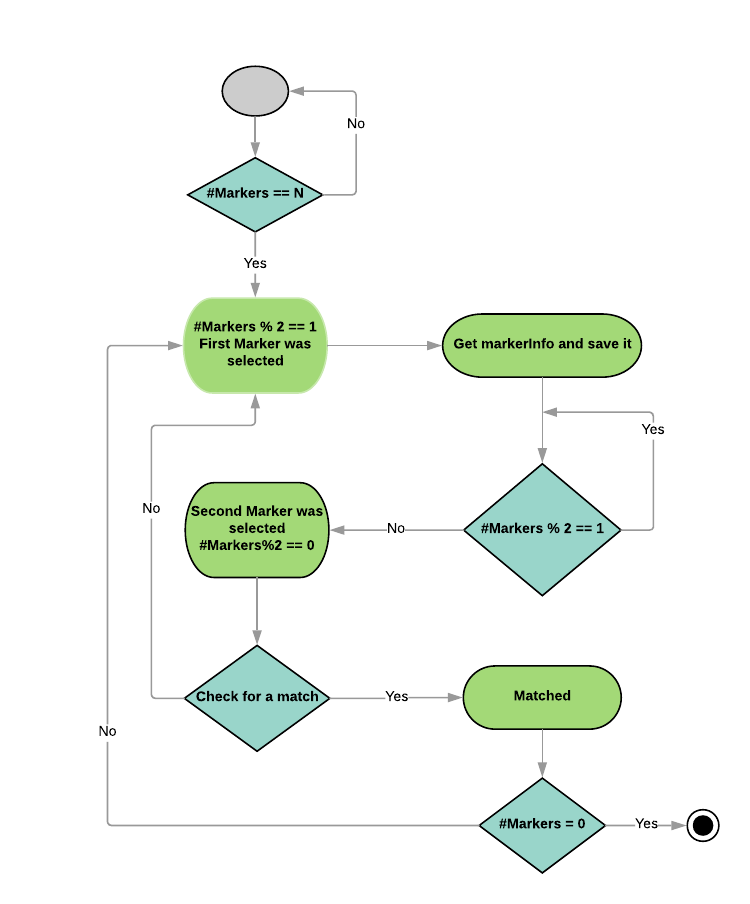}
    \caption{Activity diagram of the Finite State Machine}
    \label{rules}
\end{figure}

\subsection{Matching the Pair Game}

We created an interactive game \say{Matching the Pairs} as a collaborative task that humans can play together with the robot as a team. The goal of the game was to find all the matching pairs in a set of fiducial markers as shown in Figure \ref{pic3}. Each marker had two sides; the upper side represented the fiducial marker that was used to generate and indicate the marker ID, while the underside indicated the symbol which the marker represented, e.g. a picture of an apple, smiley face etc. The rules of the game were simple; we placed 12 markers on the table at the beginning of the game; all of the markers were downwards facing, i.e. the symbols were hidden, and the participant could only see the fiducial marker. The player received 5 points at the beginning of the game. As the game resumed, the player received 1 point for each correct match and received -1 point for each mismatch. The goal was to maintain maximum points while matching all the 12 pairs in order to win the game. 

We programmed both the Husky and the Pepper to collaboratively play the Matching the Pair game with the users. 
Firstly, the robot informed the user about the game rules and the robot’s specific functionality and reliability (error-rate) while supporting the participant with the game.


We created two conditions for both robots through varying the robot's reliability in providing help while giving instructions to the users. The robot provides help as having either a 3\% error rate or a 50\% error rate and being either human-like or machine-like in appearance to prime the participant's expectations. It is significant to note that results from a pre-test confirmed that husky was deemed machine-like and pepper was perceived more human-like. Similarly, the robot's 3\% error-rate was also judged as low error-rate and 50\% error-rate was judged as high error rate by the participants before the study.


The game was implemented based on a Finite State Machine (FSM). The algorithm and the rules for the interaction game are described in the activity diagram as shown in figure \ref{rules}. The user and the robot sat facing each other with twelve-markers placed between them (see Figure\ref{fig:setup}).  The user chooses a marker. The marker information was saved including the symbol which it represents with the use of Augmented Reality (AR) functionality; to enable the robot to track the markers. The user later had the following two options regarding the selection of the second marker:

\begin{itemize}
\item	Ask the robot for help in choosing the other matching pair by saying \say{help me}. In this case, Pepper/Husky recognizes that the participant wants help via the use of the speech recognition engine and responds accordingly. Pepper later responds verbally, while Husky responds visually i.e. displays the results on the screen.
\item Choose the marker without asking the robot for help. 
\end{itemize}



\subsection{Online System to measure Cognitive Load}

\begin{figure}
\centering
\includegraphics[width=0.45\textwidth]{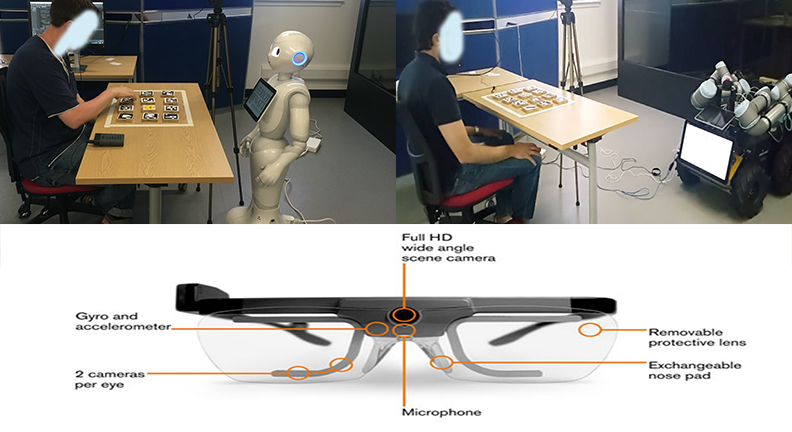}
\caption{System setup: The Tobii Glasses 2 are given to each of our participants. Our software records their pupil diameter during the interaction game with the robot. When the cognitive load is perceived as high this can be observed by the experiment as it is displayed on a screen during the interaction. }
\label{pic3}
\end{figure}

Our system connects the Tobii Pro 2 eye-tracking glasses using the Tobii Pro Glasses 2 python controller \cite{TobiiProGlasses2} library over wifi with the computer (see Figure\ref{pic3}). The new system provides information on the cognitive load of the participant at 17 fps. After a very simple calibration phase that takes advantage of the pupillar light reflex \cite{kun2012exploring}, the system is able to track the user's cognitive load, based on an empirically set threshold. Our system provides a running average $\zeta$, of pupil diameter:
\vspace{-0.2cm}
$$\zeta(d)=\frac{\sum_{t=1}^N{d_t}}{N}$$
where $d_t$ is the current pupil diameter and $N$ the number of frames.  
It further provides a windowed average:
\vspace{-0.1cm}
$$\zeta(d)=\frac{\sum_{t=1}^{15}{d_t}}{15}$$
where the average is calculated based on the past 15 frames only.
Furthermore, our system provides a running peak estimation where we assume that, when the size of the pupil is larger than  $70\%$ of the maximum the cognitive load is high.

\section{Research Method}

Our research explores two different aspects. Firstly,  we focused on the relationship between cognitive load and human's trust perception during human-robot collaboration. Secondly, 
we investigated the relationship between trust and anthropomorphism. More concretely, we explored how human's trust perception is impacted during the interaction with a human-like vs. a machine-like robot having a high vs. low error rate. 
Keeping these aforementioned aspects in mind, the following hypotheses were derived based on the literature on Human-Automation \cite{b11,b12} and on trust in HRI \cite{b3}:

\noindent\textbf{H1a:} Cognitive load is expected to predict participants' level of trust after HRI. 

\noindent\textbf{H1b:} The lower participants' cognitive load during HRI, the more trust towards the robot they will indicate after HRI.

\noindent\textbf{H2a:} Participants' cognitive load will be lower after the interaction with a human-like (vs. machine-like) robot. 

\noindent\textbf{H2b:} The impact of robot type on cognitive load predicted in H2a will be more pronounced after the interaction with a robot with a low (vs. high) error rate.

\noindent\textbf{H3a:}  Participants’ trust towards the robot will be significantly  higher  after the interaction with a human-like (vs. machine-like) robot. 

\noindent\textbf{H3b:}  The  impact  of  robot  type  on  participants’  trust towards the robot  predicted in H3a will be more pronounced after the interaction with a robot with a low (vs. high) error rate.

\vspace{3mm}
\noindent In the following hypotheses, pre- and post HRI measures of trust within subjects are considered: \vspace{3mm}

\noindent Participants' individual trust level will be more increased after the interaction with a

\noindent \textbf{H4a:} human-like (vs. machine-like) robot.

\noindent \textbf{H4b:} a robot with a low (vs. a high) error rate.


\subsection{Participants}

We conducted our between-subject study with 26 adults (8 females, and 18 males) between 25 and 46 years with a mean age of 30.46 and standard deviation of 6.50 interacting with either of the robots in either of 2 modes (high or low error rate). 
With 5 participants interacting with the Pepper on low error rate, 8 participants interacting with the Pepper high error rate, 5 participants interacting with the Husky low error rate and 8 participants interacting with the Husky high error rate. 

It is important to note that our study had 40 participants in total, however, due to difficulties in collecting data for cognitive load through Tobii glasses for some participants, we are reporting results of the 26 adults to understand the relationship between trust and cognitive load. However, for the trust perception mainly comparing the pre- and post-test results, we report results for all the 40 participants as we did not have missing data for the participants' trust during the pre- and post-test. The conditions were divided as: 10 participants interacting with the Pepper low error rate, 10 participants interacting with the Pepper high error rate, 10 participants interacting with the Husky low error rate and 10 participants interacting with the Husky high error rate.


\subsection{Procedure}

We conducted our study in four different steps: Firstly, an information sheet was handed to each participant with information regarding the robot they would be interacting with, along with an image of that robot. A written and approved consent form was obtained from each participant prior to the experiment. 

Secondly, each participant was asked to fill in an adaptation of the Godspeed questionnaire \cite{b19}, containing additional questions related to the experiment on their trust perception of the robot. 

Thirdly, each participant played the matching the pairs game with either the Husky or the Pepper robot having one of two different error-rate conditions (3\% and 50\%) in terms of providing help during the game. It is should be noted that participants were assigned randomly to one of the experimental conditions. Once the game began, both Pepper/Husky indicated the rules of the game, and provided guidance to the participants throughout the game. Low and high error conditions were incorporated on the robots with the occurrence of mistakes $3\%$ (low) and  $50\%$ (high) of the time throughout the game. The game finished once all pairs were found.

Lastly,  each participant completed the same questionnaire as in the second step. The questionnaires pre- and post– were compared using a paired and independent sample t-test.


\subsection{Setup and Materials}

\begin{figure}
    \centering
    \includegraphics[width=0.4\textwidth]{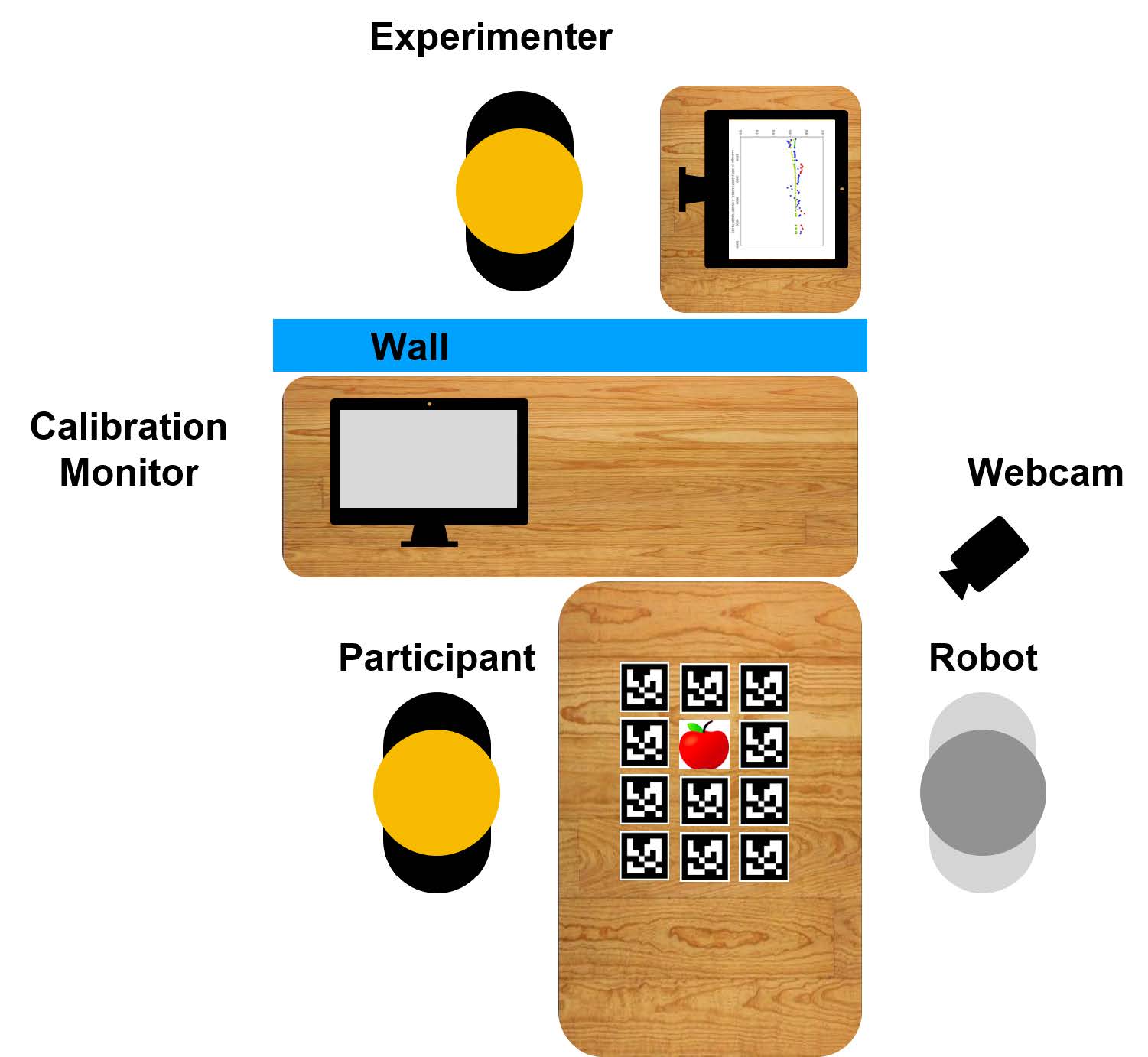}
    \caption{Overview of the experimental setup.}
    \label{fig:setup}
\end{figure}

We conducted our study in a quiet room that was divided into two parts with a divider as shown in Figure \ref{fig:setup}. On the one side, the game was placed on the table and either the Husky or the Pepper robot used in each case were placed across the table where the task was performed. The participant was seated in front of the robot. On the other side, another table was placed on which a monitor was situated for the purposes of calibrating the eye tracking equipment. The actions throughout the duration of the task were tracked with the use of AR marker technology and a webcam. Every component of the experiment was controlled from a computer operated from a concealed position behind a blind wall, by the experimenters. It is important to note that we controlled the lighting conditions to maintain the quality of the eye tracking data.

The study had two different phases that were realized in an identical fashion. For the first part, Pepper was used, along with its voice recognition capabilities for the purposes of the study. In the second part, the non-anthropomorphic robot, Husky was used. Both robots acted as optional assistants for the participants during the task they were asked to perform. Due to technical difficulties with Google speech recognition, a Wizard-of-OZ approach was used with the Husky during the experiments. Nonetheless, the same text based and speech based sentences where produced by the Husky and the Pepper robot. The Husky robot displayed the text on the screen while the Pepper robot uttered the sentences.

\subsection{Measurements}

To measure cognitive load during the HRI, we considered the number of peaks per participant as measured by the Tobii Pro 2 eye-tracking glasses. It is important to note that the duration of the game varied among the participants. However, we did not find the need to normalize the number of peaks according to the duration of gameplay. The rationale for this lies in the process of our data analysis as we mainly studied the relationship of subjective measurement of user trust and Cognitive load (number of peaks). This means the more time they spent, the more or less cognitive load they experienced is not relevant for our analysis.  
In addition, we subjectively measured trust perception on a scale from 1 to 5 through updating Godspeed questionnaires according to our study.

\section{Results}

To test the hypothesis H1a that cognitive load would predict participants' level of trust after HRI (H1a), a simple linear regression was conducted with the number of peaks as an indicator of cognitive load during HRI as a predictor of participants' level of trust towards the robot after HRI. Furthermore, a Pearson correlation between the number of peaks and participants' trust ratings after HRI was performed in order to test our prediction that the lower participants' cognitive load during HRI, the more trust towards the robot they would indicate after HRI (H1b).
In line with H1a, the number of peaks turned out to be a statistically significant predictor of participants' trust level after having interacted with the robot. The prediction model was statistically significant, \textit{F}(1,24) = 4.82, \textit{p} = .038, and accounted for approximately 17$\%$ of the variance of participants' trust ratings after HRI, $\textit{R}^2$ = .167, adjusted $\textit{R}^2$ = .132,  $\beta$ = -.409.
In line with H1b, the lower participants' cognitive load during HRI, the more trust they showed towards the robot after HRI \textit{r}(24) = -.41,  \textit{p} = .019.

\begin{figure}
    \centering
    \includegraphics[width=0.45\textwidth]{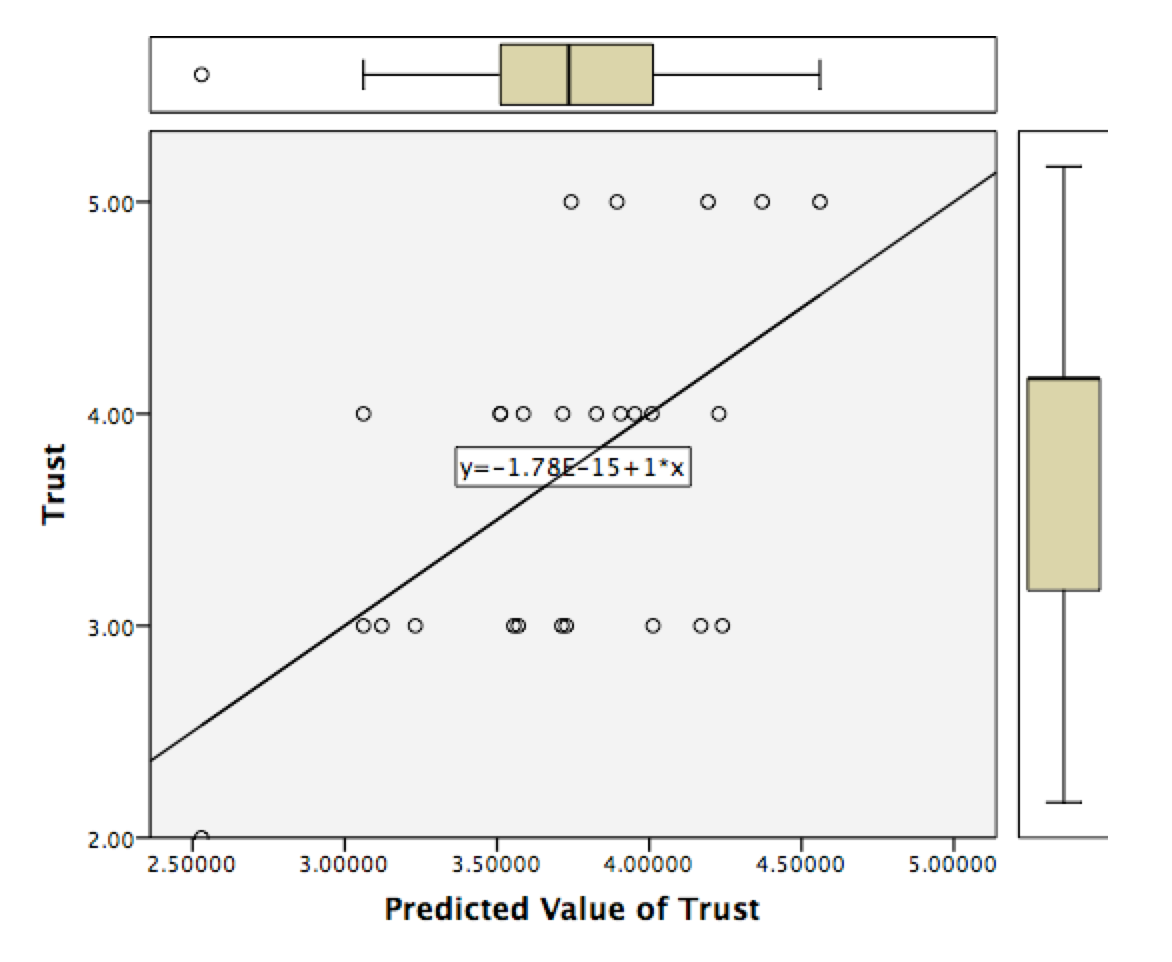}
    \caption{A linear relationship between trust and the predicted value of trust by our linear regression model.}
    \label{regression}
\end{figure}

To test our predictions that cognitive load (H2a) would be lower during HRI and participants' level of trust towards the robot (H3a) would be higher after HRI with a human-like (vs. a machine-like) robot and that the impacts of robot type on cognitive load (H2a) and trust ratings (H3a) would be even more pronounced after having interacted with a robot with a low (vs. a high) error rate (H2b, H3b), a multiple analysis of variance (MANOVA) was conducted with robot type (human-like vs. machine-like) and robot error rate (low vs. high) as factors and the number of peaks indicating participants' cognitive load and their trust towards the robot after HRI as dependent measures. Contrasting our predictions, there was neither a statistically significant main impact of robot type on participants' cognitive load during HRI, \textit{F}(1,22) = 0.99, \textit{p} = .331, $\eta_{\text{p}}^{2}$ = .043, nor on participants' trust ratings after HRI, \textit{F}(1,22) = 0.04, \textit{p} = .838, $\eta_{\text{p}}^{2}$ = .002. There were no statistically significant main impacts of error rate neither on cognitive load, \textit{F}(1,22) = 0.74, \textit{p} = .399, $\eta_{\text{p}}^{2}$ = .003, nor on participants' trust level after HRI, \textit{F}(1,22) = 2.48, \textit{p} = .130, $\eta_{\text{p}}^{2}$ = .101. Furthermore,
there was no statistically significant interaction impact between robot type and robot error rate on cognitive load, \textit{F}(1,22) = 0.79, \textit{p} = .383, $\eta_{\text{p}}^{2}$ = .035. However, the interaction between robot type and robot error rate on trust ratings after HRI was statistically significant, \textit{F}(1,22) = 4.83, \textit{p} = .039, $\eta_{\text{p}}^{2}$ = .180.

To illustrate, when robot error rate was not considered, means were seemingly in line with our predictions. That is, participants' level of cognitive load was the same after the interaction with a human-like Pepper robot (\textit{M} = 110.00, \textit{SD} = 50.10) and after the interaction with a machine-like Husky robot (\textit{M} = 90.38, \textit{SD} = 55.37). Likewise, trust ratings did not differ after having interacted with a Pepper robot (\textit{M} = 3.77, \textit{SD} = 0.73) and after having interacted with a Husky robot (\textit{M} = 3.69, \textit{SD} = 0.95). Interestingly, considering robot error rate revealed that a low error rate Pepper robot evoked relatively high levels of cognitive load (see Fig. 5), while a low error rate Husky robot evoked relatively high levels of trust after HRI (see Fig. 6). This latter finding is reflected in the statistically significant interaction impact between robot type and robot error rate on participants' trust ratings after having interacted with the robot. This interaction however, is contrasting our predictions a human-like robot with a low error rate to evoke higher levels of trust towards the robot after HRI (H3a, H3b).

\begin{figure}
    \centering
    \includegraphics[width=0.45\textwidth]{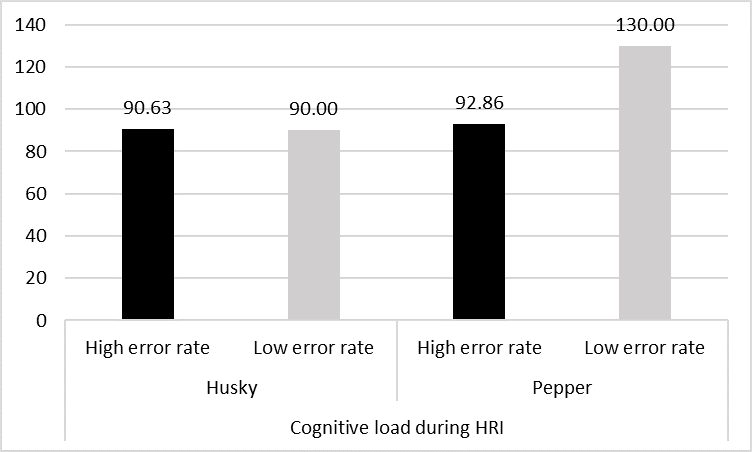}
    \caption{Participants' mean levels of cognitive load during HRI.}
    \label{Descriptives}
\end{figure}

\begin{figure}
    \centering
    \includegraphics[width=0.45\textwidth]{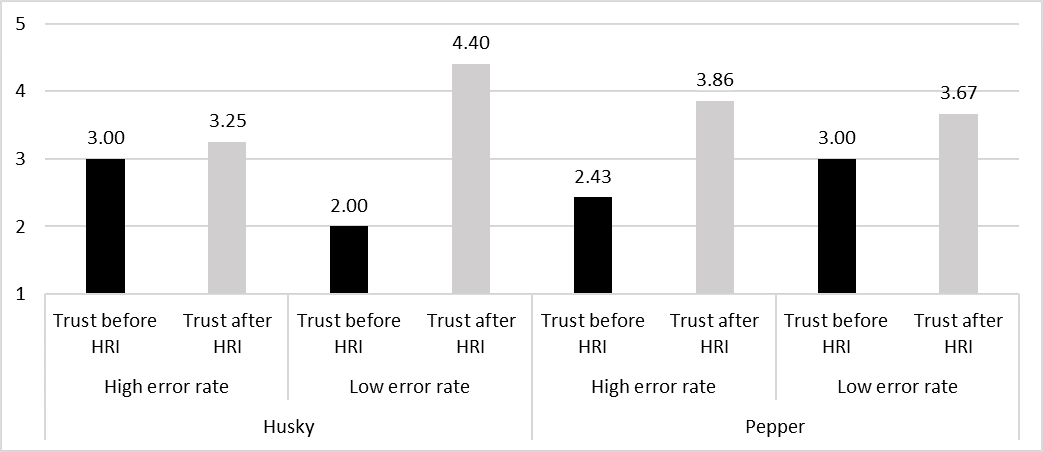}
    \caption{Participants' mean levels of trust towards the robot before and after HRI.}
    \label{Descriptives}
\end{figure}

To test our predictions that participants' individual trust level would be more increased after having interacted with a human-like Pepper robot (H4a) than after having interacted with a machine-like Husky robot and that the impact of robot type on participants' trust towards the robot after HRI would be more pronounced if the robot they had interacted with had a low (vs. a high) error rate (H4b), a repeated measures MANOVA was conducted with robot type (human-like vs. machine-like) and robot error rate (low vs. high) as factors and participants' levels of trust before HRI and after HRI as dependent variables. 

There was a statistically significant main impact of trust, sphericity assumed: \textit{F}(1,22) = 24.59, \textit{p} $<$ .001, $\eta_{\text{p}}^{2}$ = .528. That is, participants' trust towards the robot was statistically significantly higher after HRI
than before HRI was initiated (see Fig. 6).
However, contrasting our predictions there was neither a statistically significant interaction between robot type and participants' trust ratings before and after HRI took place (H4a), sphericity assumed: \textit{F}(1,22) = 0.34, \textit{p} = .568, $\eta_{\text{p}}^{2}$ = .015. More precisely, this indicates that participants' initial level of trust towards the robot was the same independent which robot type they interacted with at a later point.
However, it also indicates that contrasting to H4a, participants' trust ratings were also on a similar level after having interacted with a human-like Pepper robot
as after having interacted with a machine-like Husky robot (see Fig. 6).
Furthermore, there was no statistically interaction impact between participants' trust ratings before and after having interacted with the robot and robot error rate (H4b), sphericity assumed, \textit{F}(1,22) = 2.10, \textit{p} = .161, $\eta_{\text{p}}^{2}$ = .087. That is, participants indicated similar levels of trust before and after having interacted with a robot with a low error rate and a robot with a high error rate.
Remarkably, the interaction between participants' trust ratings before and their trust ratings after having interacted with a robot with robot type and robot error rate was statistically significant, sphericity assumed, \textit{F}(1,22) = 9.26, \textit{p} = .006, $\eta_{\text{p}}^{2}$ = .296. It needs to be referred to participants' mean scores of trust before and after HRI took place to explain this triple interaction. Comparing mean ratings on trust before and after HRI was initiated, it becomes apparent that participants' trust levels increased the most after having interacted with a low error rate Husky robot and a high error rate Pepper robot (see Fig. 6).

\section{Discussion}


We see that there was an inversely proportional relation between trust and cognitive load. We also found that cognitive load can predict participant's ratings of trust. We understand that our results are in line with the conclusions in the human-automation research where it was observed that the trust ratings of participants declined under situation demanding high cognitive load \cite{b12}.  In addition, we also found that there was a statistically significant interaction impact between robot type and error-rate on trust ratings. 
It is also notable to know that this finding is also in line with the past findings reported in human-automation use literature. It has shown that the participant's trust ratings increase when the system maintains reliability \cite{b13}. We understand that the robot with low error-rate may have resulted in this impact for both Husky and Pepper robot. 
This can also be reflected in the mean-values of participants trust rating after HRI. 

We did not see the main impact of robot type and error-rate on cognitive load. We conjecture that this could be due to the number of questions asked by the participant in the low- or high- error rate condition. 
This suggests that participants were not keen on asking for robot's assistance in the situation demanding higher cognitive load during the game \cite{minadakis2018using}. Primarily, This may be related to the error-rate condition they were in as we also find this impact as a part of our results. We also speculate that this is directly related to the limited number of participants as we did witness a higher mean value for the cognitive load for the Pepper robot in comparison with the Husky robot in the case of low error-rate. This directs and calls for a deeper investigation of the relationship between the two variables, in particular, with a higher number of participants. It is also interesting to note that this impact may also be related to the results of the studies performed in the human-automation use \cite{b6}. 

We also understand that there may be a correlation with the age of the participants and the cognitive load that we have not investigated yet. Due to using pupil diameter as a measure for cognitive load, this might be dependent on age, as the reaction time of the pupil changes during aging \cite{van2004memory}. Nevertheless, we found relatively stable results in the detection of cognitive load in the pupil diameter and therefore we believe it is a good real-time measure for the cognitive load. Other measures of cognitive load are of interest for us as well, e.g. verbal features pitch, volume or velocity which we have recently investigated here \cite{lopes2018symptoms}.  Clearly, we need to collect more data to establish that results presented are robust with different participants and that we can exchange task easily. So far our system shows a promising way of detecting cognitive load in HRI, but further evaluation and data collection are needed. Our results also find a correlation between cognitive load and trust while controlling the type of robot. This also refers to further exploration of how anthropomorphizing can impact the relationship between the two variables.



We see that there was an increase of participants' trust rating after HRI and understand that our findings are in line with the previous findings, where, the trust ratings were also significantly higher after HRI during a  card game \cite{correia2016just}. We did not see a significant difference in participant's trust rating after HRI with either Husky or Pepper and also with Husky and Pepper having low- and high- error rates. We conjecture this could be again due to the limited number of participants as the past literature has suggested a positive relationship was established between robot anthropomorphism and trust; the more anthropomorphic the robot, the more it can be trusted. 
Additionally, one of the research studies has shown that participants trust rating was significantly higher for a more reliable robot \cite{Salem:2015:YTR:2696454.2696497}.  Moreover, it could be due to participants asking a fewer number of questions under one condition as compared to others. More specifically, it is worth noting that the outcome of interacting with a robot is highly dependent on the human involved in the interaction, that is, some participants were utterly reliant on the robot, others used it as a consultant, whilst others did not depend on the robot at all. This individuality and context specifically emphasize the demand for further research be undertaken to continually develop our understanding of HRI.

One of the notable findings was that we found a triple interaction impact between robot type, error rate, and the pre- and post- measures on trust. It was fascinating to see that in the low error-rate condition, participants trusted Husky more than Pepper robot. On the contrary, participants trusted Pepper more than Husky Robot in the high error-rate condition. As highlighted in the literature, trust is a multidimensional construct that is perceived in three different dimensions: impactive, cognitive and behavioral trust \cite{johnson2005cognitive}. In HRI, a robot's performance (reliability) derives cognitive trust while users’ perception of the robot’s motives is attributed to impactive trust \cite{b3}. A group of researchers \cite{bernotat2019fe} 
recently explored the relationships between the constructs of trust and observed that participants generally indicated more cognitive trust than impactive trust towards a male vs. female human-like robot. Their findings also highlighted that participants attribute stereotypes towards robot based on their body shape. These findings help in explaining our results as Husky is an industrial robot and participants may perceive it to be more trustworthy in case of higher reliability suggesting to have may be more cognitive trust as compared to Pepper. Similarly, Pepper although less reliable was deemed more trustworthy because participants may not expect industrial robots to make mistakes as they can harm humans in this case. In summary, we understand that this triple impact is very interesting and demands more experimentation in the future research.



\section{Conclusions}

In this paper, we presented an exploratory study to understand the relationship between human cognitive load, trust, and anthropomorphism during a Human-Robot Interaction. To understand the relationship, we created a \say{Matching the Pair} game that was designed to enable humans to play the game collaboratively with robots. The idea was to understand if humans trust to work with robots as teammates during the game under situations demanding high cognitive load to complete a task. Additionally, we also exploited the aspect of anthropomorphism through enabling humans to either interact with a humanoid (Pepper) or a non-humanoid (Husky) robot. We found that there was an inversely proportional relationship between trust and cognitive load, suggesting that as the amount of cognitive load increased in the participants, their level of trust decreased. Additionally, we also found that participant's perceived the Pepper robot to be more trust worthy in comparison with the Husky robot through comparing our pre- and post-test questionnaire results.

\section{Limitation and Future Work}

It is to be noted that our online system to compute number of peaks (maximum cognitive load) is limited to the specific lighting conditions. We were limited to the number of participants for this study. In particular, we did not find enough participants to run analysis to account of age as the reaction time of the pupil varies during aging. 
Therefore, in the future, we intend to further understand the relationship between trust and cognitive load through conducting similar studies with a higher number participants and with equally balanced genders. 

\bibliographystyle{aaai}
\bibliography{pupilar}

\end{document}